\documentclass[preprint]{elsarticle}

\usepackage{graphicx}  
\usepackage{amsmath}
\usepackage{enumerate}
\usepackage{epstopdf}

\newcommand{\acrofig}{FIG.}

\begin{document}

\title{Current-driven domain wall dynamics in ferrimagnets: micromagnetic approach and collective coordinates model} 

\author[1]{Eduardo Mart\'{\i}nez
}\ead{edumartinez@usal.es}
\author[1]{{V\'{\i}ctor} Raposo
}\ead{victor@usal.es}
\author[2]{\'{O}scar Alejos
\corref{cor1}}\ead{oscar.alejos@uva.es}

\cortext[cor1]{Corresponding author}
\address[1]{Dpto. F\'{\i}sica Aplicada, University of Salamanca, 37008 Salamanca, Spain}
\address[2]{Dpto. Electricidad y Electr\'{o}nica, University of Valladolid, 47011 Valladolid, Spain}

\begin{abstract}

Theoretical studies dealing with current-driven domain wall dynamics in ferrimagnetic alloys and, by extension, other antiferromagnetically coupled systems as some multilayers, are here presented. The analysis has been made by means of micromagnetic simulations that consider these systems as constituted by two subsystems coupled in terms of an additional exchange interlacing them. Both subsystems differ in their respective gyromagnetic ratios and temperature dependence. Other interactions, as for example anisotropic exchange or spin-orbit torques, can be accounted for differently within each subsystem according to the physical structure. Micromagnetic simulations are also endorsed by means of a collective coordinates model which, in contrast with some previous approaches to these antiferromagnetically coupled systems, based on effective parameters, also considers them as formed by two coupled subsystems with experimentally definite parameters. 
Both simulations and the collective model reinforce the angular moment compensation argument as accountable for the linear increase with current of domain wall velocities in these alloys at a certain temperature or composition. Importantly, the proposed approach by means of two coupled subsystems permits to infer relevant results in the development of future experimental setups that are unattainable by means of effective models.
\end{abstract}

\maketitle

\section{Introduction}
The development of racetrack memories\cite{Parkin:08,Kim:10} has attracted much interest in the recent times. Many efforts have been addressed in that way, particularly, the finding of optimal systems allowing fast displacement of domain walls (DWs) along them. Interfacial effects such as the Dzyaloshinskii-Moriya interaction, along with the generation of spin currents through the Spin-Hall effect (SHE) constituted a major step to this target.\cite{Thiaville:12,Chen:13,Tetienne:15} The current-driven domain walls dynamics (CDDWD) in these systems are characterized by the absence of DW precessional regimes, against dynamics driven by magnetic fields or spin transfer torques.\cite{Mougin:07,Martinez:12} Nevertheless, the highest velocities reached by DWs under SHE are limited due to the reorientation of DW magnetic moments with the polarization of the spin current, so that velocity saturates as driving currents are increased.\cite{Martinez:14b} Besides, experiments demonstrate that the DW transition type determines the velocity gained by DWs when tracking curved paths. In fact, some DWs can run faster or slower than others if they pass through a strip curved section. Velocities also differ from that gained by DWs tracking straight paths.\cite{Garg:17}

Recent experimental evidence shows that current-driven DWs can reach velocities as fast as 1km$\cdot$s\textsuperscript{–-1} along strips formed by antiferromagnetically coupled bilayers.\cite{Blasing:18} A dragging mechanism, resulting in a vanishing DW tilting, has been proved to allow such fast and synchronous tracking of DWs, even along curved paths. \cite{Blasing:18, Alejos:18} Besides, it has been found under certain conditions a linear relationship between DW velocities and current magnitudes, as also occurs in the case of certain ferrimagnetic alloys.\cite{Caretta:18}

Based on these promising experimental results, we provide full micromagnetic ($\mu$M) studies dealing with CDDWD in ferrimagnets (FiMs), also extendable to other antiferromagnetically coupled systems.\cite{Alejos:18} Our $\mu$M simulations treat them as constituted by two subsystems, in particular, two sublattices in FiMs, coupled by means of an additional intersystem exchange, and differing in their gyromagnetic ratios and saturation magnetization. Some other interactions can be accounted for in distinct manners within each subsystem, depending on the considered physical characteristics. $\mu$M simulations are also backed up with the help of an extended collective coordinates model in the form of a one-dimensional model (1DM).\cite{Alejos:18} In contrast with previous approaches to these systems, based on effective parameters,\cite{Je:18} this 1DM also considers systems as formed by two coupled subsystems with experimentally definite parameters. This is rather important, since the 1DM here presented does not require diverging parameters to match the results of the experimental evidence.\cite{Kim:19} Importantly, the gyromagnetic ratio and the Gilbert damping parameter are taken as constants, contrarily to what it was suggested in the literature.\cite{Caretta:18} Realistic conditions can also be evaluated. Therefore, such approach permits to infer results not achievable from the mentioned effective models that can be of relevance in the development of future experimental setups. In particular, the work confirms the alignment of the magnetization with the direction of the electric current as responsible for the linear increase of DW terminal velocities with current. This alignment is remarkable when the angular moment compensation of sublattices occurs at a temperature close but not equal to the magnetization compensation temperature. Last but not least, the CDDWD along curved paths is found here to be dissimilar to that along straight ones, with distinct velocities depending on the DW type.


\section{Models}
\begin{figure}[th]
\centering
  \begin{tabular}{@{}cc@{}}
    (a)\includegraphics[width=30mm]{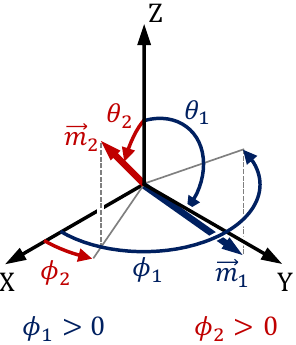} &
    (b)\includegraphics[width=47mm]{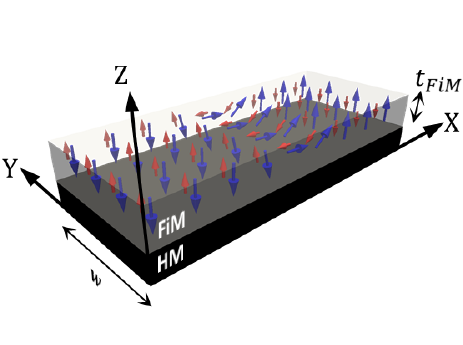}\\
    \multicolumn{2}{c}{(c)\includegraphics[width=50mm]{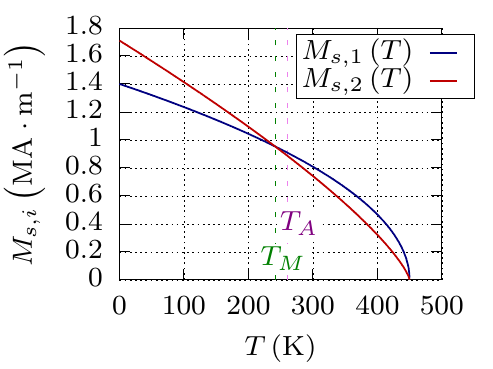}}
  \end{tabular}
  \caption{Two sublattices of respective magnetizations $M_{s,1}$ and $M_{s,2}$ constitute the ferrimagnetic strip: (a) unit vectors $\vec{m}_1$ and $\vec{m}_2$ defining the local orientation of the magnetization within each sublattice, (b) magnetic DW of N{\'e}el type amidst two domains oriented out of plane (the strip width $w$ is here shown), and (c) temperature dependence of the magnetization of each sublattice.}
  \label{Fig:01}
\end{figure}

Figure \ref{Fig:01} shows the main characteristics of the system under study. A FiM strip is deposited on top of a heavy metal (HM) layer. FiMs are considered to be formed by two sublattices of magnetizations $M_{s,1}$ and $M_{s,2}$ which are oriented along the unit vectors $\vec{m}_1$ and $\vec{m}_2$. Due to the perpendicular magnetic anisotropy, magnetizations point out of plane within the magnetic domains. Since magnetostatic interactions are much weaker than anisotropic exchange, such as the Dzyaloshinskii-Moriya interaction, DWs in the system are of N{\'e}el type. Temperature dependencies are also plotted in (c). Magnetizations decrease with temperature according to the law $M_{s,i}\left(T\right)=M_{s,i}^0\left(1-\frac{T}{T_C}\right)^{a_i}$, $i=1,2$, $M_{s,i}^0$ being the respective magnetizations at zero temperature. $T_C$ is the Curie temperature of the system, and $a_i$ are some exponents depending on the sublattice components. Importantly, both magnetizations equalize at the magnetization compensation temperature $T_M$, leading to a null net magnetization, whereas magnetizations meet the condition $\frac{M_{s,1}}{\gamma_1}=\frac{M_{s,2}}{\gamma_2}$ at the angular momentum compensation temperature $T_A$, with $\gamma_1$ and $\gamma_2$ being the respective gyromagnetic ratios for each sublattice.

\subsection{Micromagnetic modeling}
In the framework of the $\mu$M model, the magnetization dynamics of a single ferromagnetic strip is described by the Landau-Lifshitz-Gilbert (LLG) equation.\cite{Haazen:13,Zhang:04} This idea can be generalized to study the magnetization dynamics in the FiM by considering two LLG equations, one for each sublattice, as:
\begin{align}
\begin{split}
\frac{d\vec{m} _i}{dt}=-\gamma_i\ \vec{m}_i\times \vec{H}_{eff,i}&+\alpha_i\vec{m}_i\times\frac{d\vec{m}_i}{dt}+\\
&+\vec\tau_{STT,i}+\vec\tau_{SOT,i}\text{,}
\end{split}\label{eqn:01}
\end{align}
with $i=1,2$, and $\alpha_i$ being the respective Gilbert constants of each sublattice. The effective fields $\vec{H}_{eff,i}$ summarize all interactions within the system, including all intra-sublattice interactions together with an interaction term between both sublattices, i.e.:
\begin{equation}
\vec{H}_{eff,i}=\vec{H}_{ext}+\vec{H}_{dmg,i}+\vec{H}_{ani,i}+\vec{H}_{exch,i}+\vec{H}_{D,i}\text{,}
\end{equation}
with $\vec{H}_{ext}$, $\vec{H}_{dmg,i}$, $\vec{H}_{ani,i}$, $\vec{H}_{exch,i}$ and $\vec{H}_{D,i}$ being respectively the external field, with components $\left(H_x,H_y,H_z\right)$, the magnetostatic (demagnetizing) field, the anisotropy field, with out-of-plane and in-plane components given by the effective anisotropy constants $K_{eff,i}$ and $K_{sh,i}$, the isotropic exchange field and the anisotropic exchange field. The latter is responsible for the chiral nature of magnetic textures, and is determined by certain constants $D_i$. As for the isotropic exchange field, it can be reduced on first approach to a sum of the following terms:\cite{Ma:16}
\begin{equation}
\vec{H}_{exch,i}=\frac{2A_i}{\mu_0M_{s,i}}\nabla^2\vec{m}_i+\frac{2A_{ij}}{\mu_0M_{s,i}}\nabla^2\vec{m}_j+\frac{B_{12}}{\mu_0M_{s,i}}\vec{m}_j\text{,}
\end{equation}
with $j=1,2$, and $j\neq i$. The first term here represents an intra-sublattice exchange field, given by the exchange stiffness $A_i$, and related to the local variation of the sublattice magnetization, whereas the latter is an inter-sublattice exchange field due to the misalignment of both sublattices, which is worked out from an energy density term in the form $\epsilon_{12}=-B_{12}{\vec{m}}_1\cdot{\vec{m}}_2$, so that $B_{12}>0$ ($<0$) forces the antiparallel (parallel) alignment of the sublattices. The middle term in this exchange field stands for an additional inter-sublattice exchange due to the local variations of the magnetization in one of the sublattices. This term has been implemented into our homemade $\mu$M code.\cite{Alejos:18} However, no relevant differences in the case here presented have been found between the results either including or ignoring such an additional term. Finally, $\vec\tau_{STT,i}$ and $\vec\tau_{SOT,i}$ stand for the torques due to spin polarized currents, i.e., the spin transfer torques\cite{Haazen:13} and the spin-orbit torques\cite{Zhang:04}, respectively. The former terms consist of adiabatic interactions $\vec\tau_{A,i}$, defined by certain values $u_i$, and their non-adiabatic counterparts $\vec\tau_{NA,i}$, related to these by the factors $\beta_i$, so that:
\begin{equation}
\begin{split}
\vec\tau_{STT,i}&=\vec\tau_{A,i}+\vec\tau_{NA,i}=\\
&=u_i\left(\vec{u}_J\nabla\right)\vec{m}_i-\beta_iu_i\vec{m}_i\times\left(\vec{u}_J\nabla\right)\vec{m}_i\text{,}
\end{split}
\end{equation}
where $\vec{u}_J$ represents the direction of the current density. The latter terms can be split into field-like torques with effective fields $H_{FL,i}$, and Slonczewskii-like torques with effective fields $H_{SL,i}$, both to be further on defined. In fact,
\begin{equation}
\begin{split}
\vec\tau_{SOT,i}&=\vec\tau_{FL,i}+\vec\tau_{SL,i}=\\
&=-\gamma_iH_{FL,i}\vec{m}_i\times\sigma-\gamma_iH_{SL,i}\vec{m}_i\times\left(\vec{m}_i\times\sigma\right)\text{,}
\end{split}
\end{equation}
where $\vec\sigma$ is the unit vector along the direction of the polarization of the spin current generated by SHE in the HM, that is, $\vec\sigma=\vec{u}_J\times\vec{u}_z$, since $\vec{u}_z$ defines the HM/FiM interface. $\vec{\sigma}=-\vec{u}_y$ for a longitudinal current so that $\vec{u}_J=\vec{u}_x$.\cite{Haazen:13,Ryu:13,Emori:13}

\subsection{Collective coordinates model}
Starting from (\ref{eqn:01}), and with the use of variational principles, a 1DM can be derived, leading to a set of equations that involve the instantaneous position $q$ of a DW in the system and orientations $\psi_i$ of the in-plane components of the magnetization of each sublattice with respect to the longitudinal axis of the strip. A note must be made at this time. The model could have considered different positions $q_1$ and $q_2$ for the magnetic transitions in each subnet. However, both a brief energetic argument and the study by means of micromagnetic simulations, allows inferring that the transitions occupy identical positions, i.e., $q=q_1=q_2$. Thus, the ansatz $\theta_i=2\arctan{\text{e}^{Q_i\frac{x-q}{\Delta}}}$ and $\phi_i=\psi_i$ is to be used, $\theta_i$ and $\phi_i$ defining the local orientation of the magnetization in each sublattice, as shown in \acrofig\ref{Fig:01}(a), and $\Delta$ standing for the DW width. $Q_i=\pm 1$ indicates up-down ($+1$) or down-up ($-1$) DW configurations within each sublattice. $Q_1=-Q_2$ for antiparallel coupling of sublattices.

The derivation of the 1DM results in the following set of expressions (more details in \ref{appendix:1DM}):

\begin{subequations}
\begin{align}
\begin{split}
\alpha_1\frac{M_{s,1}}{\gamma_1}\frac{\dot{q}}{\Delta}&+\alpha_2\frac{M_{s,2}}{\gamma_2}\frac{\dot{q}}{\Delta}+\frac{M_{s,1}}{\gamma_1}Q_1{\dot{\psi}}_1+\frac{M_{s,2}}{\gamma_2}Q_2{\dot{\psi}}_2=\\
&=Q_1M_{s,1}\left[H_z-\frac{\pi}{2}H_{SL,1}\cos{\psi_1}\right]+\\
&+Q_2M_{s,2}\left[H_z-\frac{\pi}{2}H_{SL,2}\cos{\psi_2}\right]+\\
&+\frac{M_{s,1}}{\gamma_1}\beta_1\frac{u_1}{\Delta}+\frac{M_{s,2}}{\gamma_2}\beta_2\frac{u_2}{\Delta}\text{,}
\end{split}
\end{align}
\begin{align}
\begin{split}
-Q_1\frac{\dot{q}}{\Delta}&+\alpha_1{\dot{\psi}}_1=-Q_1\frac{u_1}{\Delta}
-\gamma_1\frac{H_{k,1}}{2}\sin{\left(2\psi_1\right)}+\\
&+\gamma_1\frac{\pi}{2}Q_iH_{D,1}\sin{\psi_1}
-\gamma_1\frac{\pi}{2}H_{FL,1}\cos{\psi_1}+\\
&+\gamma_1\frac{\pi}{2}\left(H_y\cos{\psi_1}-H_x\sin{\psi_1}\right)+\\
&+\gamma_1\frac{2B_{12}}{\mu_0M_{s,1}}\sin{\left(\psi_1-\psi_2\right)}\text{,}
\end{split}
\end{align}
\begin{align}
\begin{split}
-Q_2\frac{\dot{q}}{\Delta}&+\alpha_2{\dot{\psi}}_2=-Q_2\frac{u_2}{\Delta}
-\gamma_2\frac{H_{k,2}}{2}\sin{\left(2\psi_2\right)}+\\
&+\gamma_2\frac{\pi}{2}Q_2H_{D,2}\sin{\psi_2}
-\gamma_2\frac{\pi}{2}H_{FL,2}\cos{\psi_2}+\\
&+\gamma_2\frac{\pi}{2}\left(H_y\cos{\psi_2}-H_x\sin{\psi_2}\right)+\\
&+\gamma_2\frac{2B_{12}}{\mu_0M_{s,2}}\sin{\left(\psi_2-\psi_1\right)}\text{,}
\end{split}
\end{align}
\end{subequations}
To ease notation, some values have been used in these equations, such as $H_{D,i}=\frac{D_i}{\mu_0M_{s,i}\Delta}$, and $H_{k,i}=\frac{2K_{sh,i}}{\mu_0M_s}$. The Sloncewskii-like term defined by means of the value $H_{SL,i}$ is related to the longitudinal current density $J_{HM}$ along the HM layer through $H_{SL,i}=\frac{\hbar\theta_{SH,i}J_{HM}}{2\left|e\right|\mu_0M_{s,i}t_{FiM}}$, with $\hbar$ being the reduced Plank constant, $\left|e\right|$ the absolute electron charge, $\theta_{SH,i}$ the spin-Hall angle for each sublattice and $t_{FiM}$ the thickness of the FiM.\cite{Martinez:13c} The field-like counterpart of this term can be regarded as proportional to it in a factor $k_i$, that is, $H_{FL,i}=k_iH_{SL,i}$.

\section{Results}
The CDDWD results here presented have been computed by means of $\mu$M simulations with the mentioned homemade code,\cite{Alejos:18} and the 1DM above developed. A common set of material parameters found in the literature\cite{Caretta:18} has been considered for the two sublattices ($i=1,2$): $A_i=70\frac{\text{pJ}}{\text{m}}$, $K_{eff,i}\approx K_{u,i}= 1.4\frac{\text{MJ}}{\text{m}^3}$, $K_{u,i}$ being the magnetic uniaxial anisotropy constant of the FiM sublattices, $K_{sh,i}\approx 0$, $\alpha_i = 0.02$, $D_i = 0.12\frac{\text{mJ}}{\text{m}^2}$, $\theta_{SH,i} = 0.155$, $k_i \approx 0$ and $u_i \approx 0$. The antiferromagnetic coupling between the two sublattices is accounted for by the parameter $B_{12} = -90\frac{\text{MJ}}{\text{m}^3}$.\cite{Ma:16} The gyromagnetic ratios ($\gamma_i = \frac{g_i\mu_B}{\hbar}$, $\mu_B$ being Bohr's magneton) are different due to distinct Land{\'e} factors: $g_1 = 2.05$ and $g_2 = 2.0$. The temperature dependence of the magnetization of each sublattice is determined by the Curie temperature $T_C = 450\text{K}$ of the FiM, and $M_{s,1}^0 = 1.4\frac{\text{MA}}{\text{m}}$ and $M_{s,2}^0 = 1.71\frac{\text{MA}}{\text{m}}$, with $a_1 = 0.5$ and $a_2 = 0.76$. According to these values, the temperature of magnetization compensation is $T_M \approx 241.5\text{K}$, and the angular momentum compensation temperature is $T_A=260\text{K}$. The dimensions of the FiM strips are $w\times t_{FiM}=256\text{nm}\times 6\text{nm}$. With these parameters, DW width is $\Delta\approx 6\text{nm}$.

The computation of $\mu$M calculations requires a previous space discretization that, in our case, has been chosen to be as small as $1\text{nm}$ cell-size.

\subsection{Transient response of domain walls in ferrimagnets}

\begin{figure}[t]
\centering
  \begin{tabular}{@{}c@{}}
    (a)\includegraphics[width=60mm]{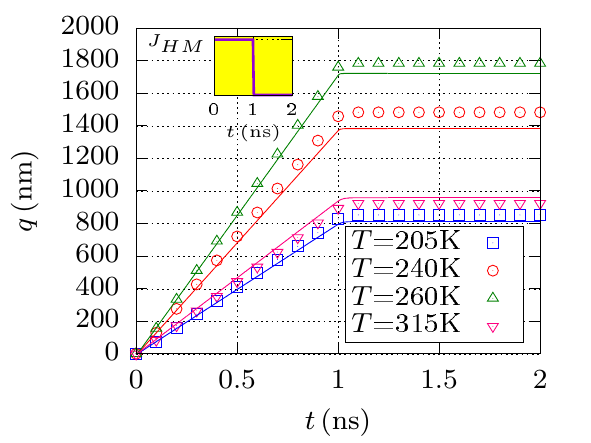} \\
    (b)\includegraphics[width=65mm]{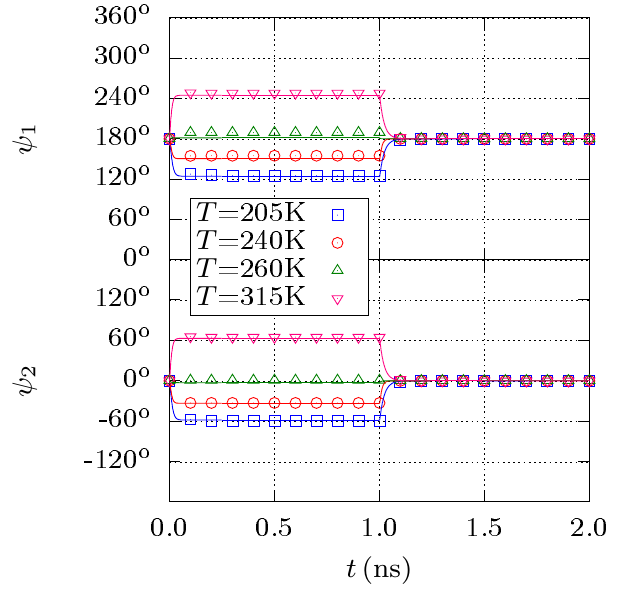}
  \end{tabular}
  \caption{Transient response of a DW under the effect of a current pulse. Pulse amplitude and duration are $J_{HM}=2\frac{\text{TA}}{\text{m}^2}$ and $t_p=1\text{ns}$. The response is obtained at four different temperatures for a) DW position $q$, and b) DW magnetization angles of both sublattices $\psi_1$, and $\psi_2$. Dots correspond to $\mu$M simulations, while continuous lines are the results drawn by the 1DM.}
  \label{Fig:02}
\end{figure}

As a first result, the transient response of a DW in a FiM straight strip under the effect of a pulse of longitudinal current $J_{HM}$ of a certain duration is presented in \acrofig\ref{Fig:02}. Graphs describe the DW displacement $q$ and the DW orientation angles $\psi_1$ and $\psi_2$ computed at four different temperatures, including magnetization compensation and angular momentum compensation temperatures ($T\approx 240K$ and $T\approx 260K$, respectively). There is a good agreement between $\mu$M and 1DM results. Both methods reveal a fast response of DWs, whose terminal velocity can be then estimated as the quotient of DW run distance and pulse duration. Importantly, the low damping considered in these simulations, as recently suggested by some authors, \cite{Kim:19} does not result in large domain wall inertia,\cite{Vogel:12, Torrejon:16} quite the contrary.

It is to note that the highest velocity is reached at the angular momentum compensation temperature ($T\approx 260K$). At this temperature, a noticeable alignment between the DW moments of both sublattices and the longitudinal current occurs (DW magnetization angles are closer to either 0\textsuperscript{o} and 180\textsuperscript{o} than at other temperatures), with a rather slight misalignment between sublattices.

\subsection{Stationary response of domain walls in ferrimagnets}
\subsubsection{Temperature dependence of CDDWD}
\begin{figure}[t]
\centering
  \begin{tabular}{@{}c@{}}
    (a)\includegraphics[width=60mm]{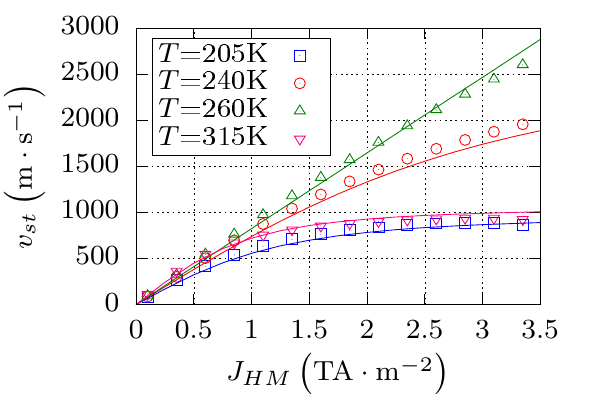} \\
    (b)\includegraphics[width=65mm]{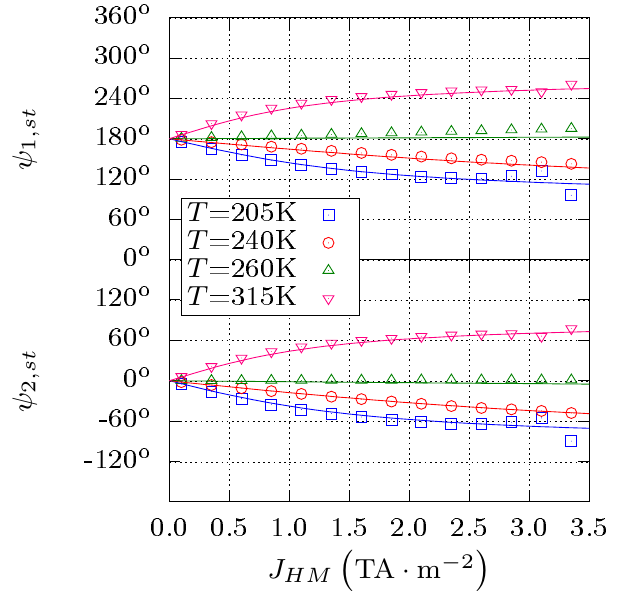}
  \end{tabular}
  \caption{Dependence on current density $J_{HM}$ of (a) DW terminal velocity $v_{st}$, and (b) DW stationary angles $\psi_{1,st}$ and $\psi_{2,st}$. Dots correspond to $\mu$M simulations, while continuous lines are the results drawn by the 1DM.}
  \label{Fig:03}
\end{figure}
\acrofig\ref{Fig:03} shows the dependencies of DW terminal velocities, $v_{st}=\dot{q}_{st}$, and stationary DW angles $\psi_{1,st}$, and $\psi_{2,st}$ with $J_{HM}$, both micromagnetically computed (dots) and calculated using the 1DM (continuous lines). Curves are obtained for different temperatures. A relevant agreement of these results and experiments\cite{Caretta:18} has been found for currents above the experimental threshold values (the effect of some imperfections in realistic samples is to be discussed further on). The rather good agreement between $\mu$M and 1DM results is also noticeable. At the compensation temperature $T_A$, $v_{st}$ increases almost linearly with current. Although the spin-polarized current promotes a slight misalignment between the magnetization of the sublattices, the linearity holds as long as sublattice internal DW moments remain closely oriented with the longitudinal direction, i.e., perpendicular to the spins of the spin-Hall current, as it can be checked also in \acrofig\ref{Fig:03}(b).  
Accordingly, a linear relationship between $v_{st}$ and $J_{HM}$ can be derived by considering DW angles completely oriented with the current direction:
\begin{equation}
v_{st}=\gamma_0\frac{\Delta}{\alpha}\frac{\pi}{2}\frac{\hbar\theta_{SH}J_{HM}}{2\left|e\right|\mu_0t_{FiM}}\frac{g_1g_2}{\left({g_2}M_{s,1}+{g_1}M_{s,2}\right)}\text{.}
\end{equation}

\begin{figure}[t]
\centering
\includegraphics[width=78mm]{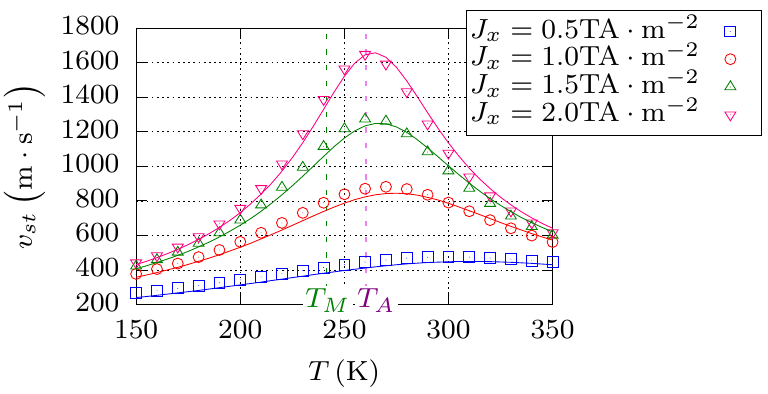} \\
\caption{Dependence of the terminal velocity $v_{st}$ with temperature for different driving currents obtained from $\mu$M simulations (dots), and the 1DM (continuous lines).}
\label{Fig:04}
\end{figure}

The general dependence of $v_{st}$ on temperature for different driving currents is plotted in \acrofig\ref{Fig:04}. Low currents (blue curve) are not sufficient to disorientate sublattice magnetizations, so that $v_{st}$ slightly varies in the considered temperature range. However, a peak around $T_A$ is observable for high currents, when the spin-Hall induced torque tends to vanish and, consequently, the spin-Hall effective field, proportional to the longitudinal internal DW moment,\cite{Thiaville:12, Emori:13, Haazen:13} is optimized.

\subsubsection{CDDWD dependence on external in-plane fields}
\begin{itemize}
\item Longitudinal fields.
\begin{figure}[t]
\centering
  \begin{tabular}{@{}c@{}}
    (a)\includegraphics[width=60mm]{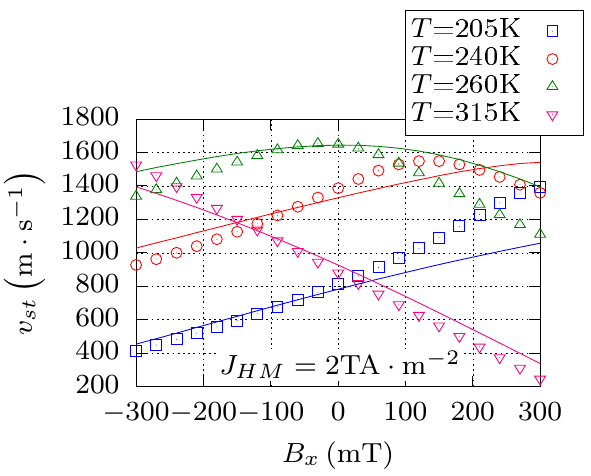} \\
    (b)\includegraphics[width=65mm]{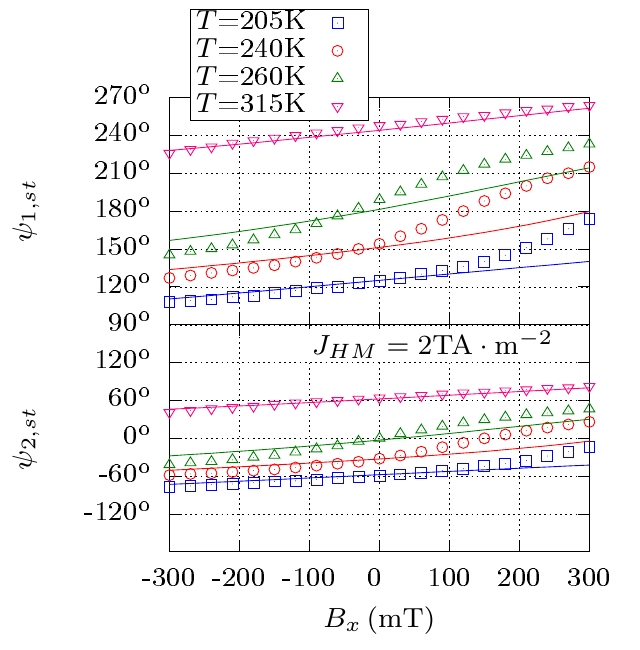}
  \end{tabular}
  \caption{Dependence on longitudinal applied field $B_x$ of (a) DW terminal speed $v_{st}$, and (b) DW stationary angles $\psi_{1,st}$ and $\psi_{2,st}$. A current density of $J_{HM}=2\frac{\text{TA}}{\text{m}^2}$ has been used. As in other graphs, dots correspond to $\mu$M simulations, while continuous lines are the results drawn by the 1DM.}
  \label{Fig:05}
\end{figure}

In order to confirm the relevance of the alignment of DW moments with current, an additional study of the effects of in-plane fields on the CDDWD has been performed. First, results for a current density of $J_{HM}=2\frac{\text{TA}}{\text{m}^2}$ and longitudinal fields $B_x$ ranging from $-300\text{mT}$ to $+300\text{mT}$ are plotted in the graphs of \acrofig\ref{Fig:05}. These results confirm that the highest speed is reached when the DW moments of both sublattices are oriented with the current direction, and this occurs when the field and the net magnetization of the system are equally oriented. In fact, since the net magnetization in the system here analyzed has opposite orientation at $315K$ and $205K$, $v_{st}$ decreases/increases with positive/negative fields (net magnetization and applied field are antiparallel/parallel) for the highest temperature and increases/decreases for the lowest one. In any case, the magnetization of each sublattice remains approximately antiparallel to each other in the whole range under study.

\item Transverse fields.
\begin{figure}[t]
\centering
  \begin{tabular}{@{}cc@{}}
    (a)\includegraphics[width=60mm]{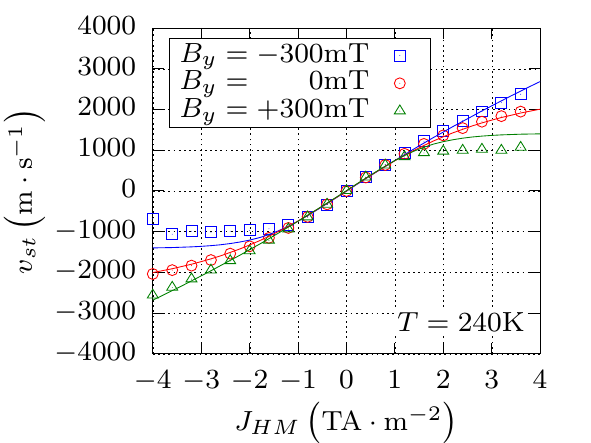} &
    (c)\includegraphics[width=60mm]{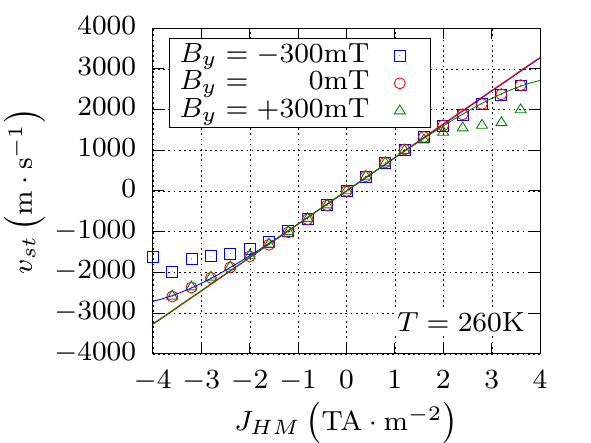} \\
    (b)\includegraphics[width=65mm]{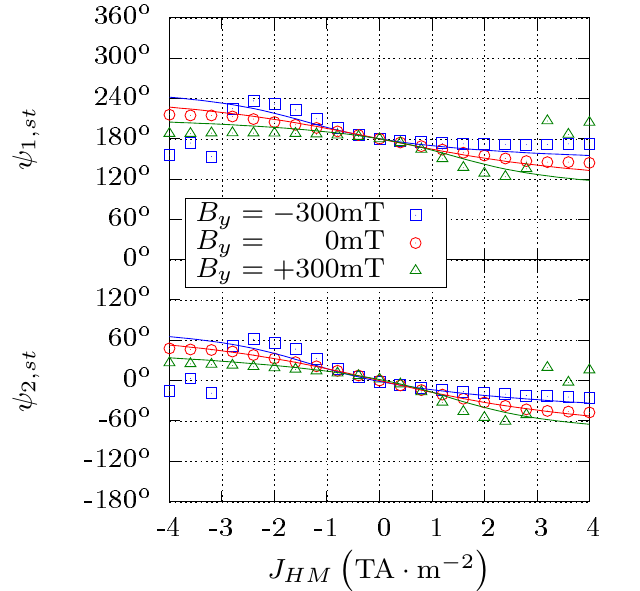} &
    (d)\includegraphics[width=65mm]{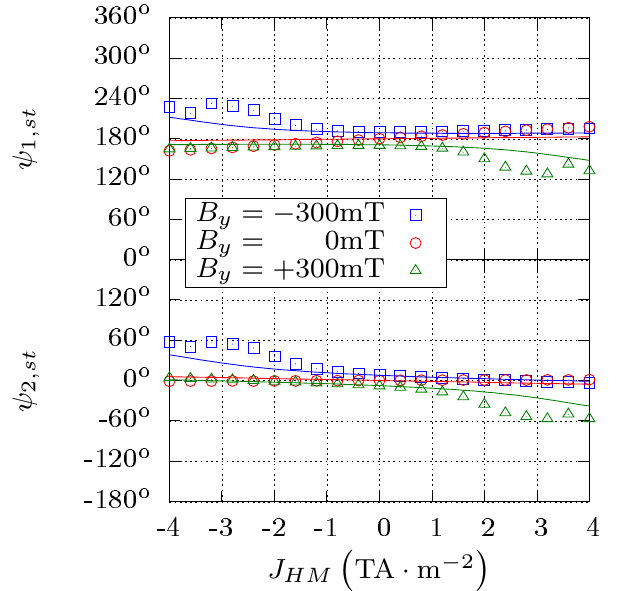}
  \end{tabular}
  \caption{Dependence on current density $J_{HM}$ with the transverse applied field $B_y$ as a parameter of (a) DW terminal velocities $v_{st}$ at $T=240\text{K}$, (b) DW stationary angles $\psi_{1,st}$ and $\psi_{2,st}$ at $T=240\text{K}$, (c) DW terminal velocities $v_{st}$ at $T=260\text{K}$, and (d) DW stationary angles $\psi_{1,st}$ and $\psi_{2,st}$ at $T=260\text{K}$. Again, dots correspond to $\mu$M simulations, while continuous lines are the results drawn by the 1DM.}
  \label{Fig:06}
\end{figure}

So as to promote the misalignment between sublattices, a transverse in-plane field $B_y$ can be applied. \acrofig\ref{Fig:06} present the corresponding results, where DW terminal velocities and stationary values of the DW angles are plotted as a function of the longitudinal current $J_{HM}$ with the transverse field as a parameter. The field and the polarization $\vec{\sigma}$ of the spin current now share the same direction. For both analyzed temperatures, an almost linear behavior is reached when both the field and $\vec{\sigma}$ are equally oriented, differently to the case when both vectors are antiparallel, and the linear behavior correspond to the cases when the DW moments of both lattices remain close to the longitudinal direction. In the absence of driving currents, the transverse in-plane field is not able to induce a continuous displacement of DWs, and a null $v_{st}$ is obtained.\end{itemize}
\subsection{Realistic samples}
\begin{figure}[t]
\centering
  \begin{tabular}{@{}cc@{}}
    (a)\includegraphics[width=50mm]{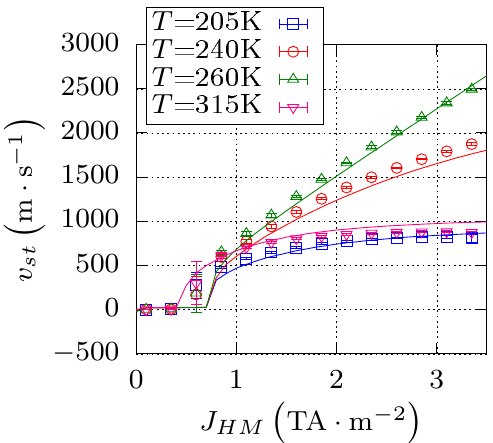} &
    (b)\includegraphics[width=50mm]{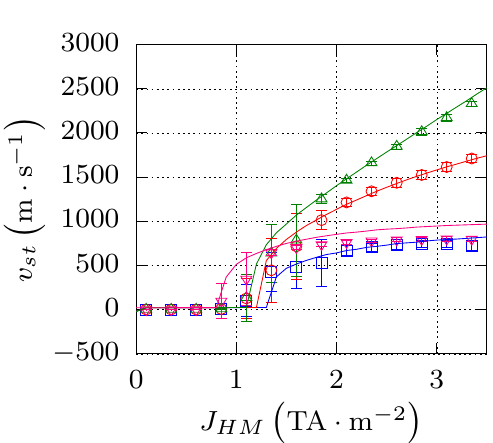}
  \end{tabular}
  \caption{Emergence of a threshold current for strips with imperfections: (a) $5\%$ anisotropy variation, and (b) $10\%$ anisotropy variation. Dots correspond to $\mu$M simulations and continuous lines to the 1DM.}
  \label{Fig:07}
\end{figure}

The effect of imperfections on the CDDWD can also be included in our analysis. A certain granularity is added to the system to mimic these imperfections. Grains of $10\text{nm}$-size have been used, characterized by an arbitrary orientation of their anisotropy axis around the out-of-plane direction. Two cases are here presented, either $5\%$ or $10\%$ variation of this orientation. Five distinct realizations have been performed for each granularity to obtain some statistically relevant results. An increasing threshold current with the degree of imperfections arise in this system (see \acrofig\ref{Fig:07}). The 1DM is also able to describe the CDDWD by using an additional periodic potential\cite{Martinez:13c} of a certain amplitude and period equal to twice the grain size. The potential amplitude used in the case of $10\%$ variation doubles that used in the $5\%$ case.

\subsection{Application to the CDDWD along curved paths}
\begin{figure}[!h]
\centering
\includegraphics[width=69mm]{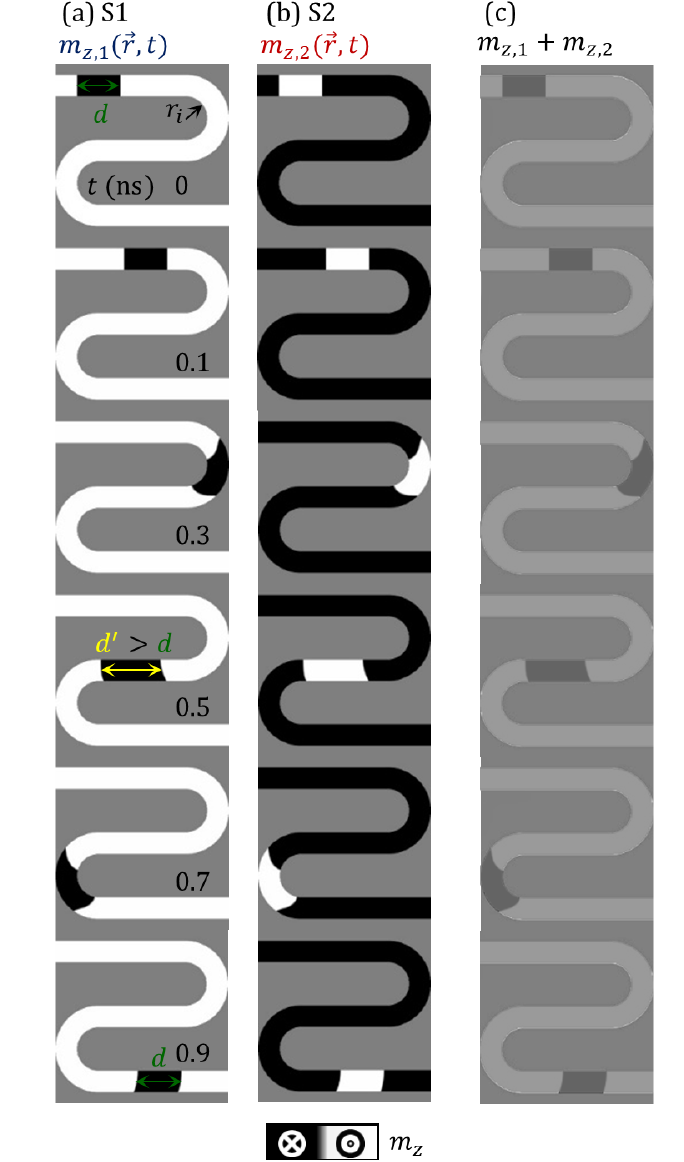} \\
\caption{$\mu$M snapshots of the CDDWD of two adjacent DWs along curved paths. The current density in the straight paths, where it is uniform, is of $J_{HM}=1.5\frac{\text{TA}}{\text{m}^2}$, and the temperature chosen is $T=T_A$. The panels represent: a) out of plane component $m_{z,1}$ of the magnetization along the first sublattice, b) out of plane component $m_{z,2}$ of the magnetization along the second sublattice, c) out of plane component of the net magnetization $m_{z,1}+m_{z,2}$ along the whole system. 
}
\label{Fig:08}
\end{figure}

Finally, we have also performed a $\mu$M analysis of the CDDWD along a strip with straight and curved parts at the most efficient case, i.e., $T=T_A$. Identical HM/FiM cross section has been adopted. Transient $\mu$M snapshots are shown in \acrofig\ref{Fig:08}. In the curved parts, the internal radius ($r_i$) is equal to the FiM strip width ($w$), i.e., $r_i = w = 256\text{nm}$. In the straight parts, DWs are very efficiently driven by the current injected along the HM without tilting. However, DWs tilt as they go through the curved parts, with similar tilting angles for all DWs of the two sublattices. We have verified that such a tilting is due to the non-uniform distribution of the current along the HM due to its curved shape. Importantly, the relative distance between adjacent DWs is not preserved after passing through the first curve. Therefore, these observations indicate the DW motion is sensitive to the curvature of the HM/FiM stacks, and further investigations are needed to correct this drawback for DW recording devices based on these systems.

\section{Conclusions}

$\mu$M simulations and the 1DM model have been proved to adequately describe the CDDWD in FiMs. In contrast with previous approaches, the procedure here presented is not based on phenomenological but on experimentally determined system parameters. This effective approach permits us, for example, to confirm that the basis of the linearity between terminal velocities and current densities at $T=T_A$ in FiMs is the fact that internal DW moments of the sublattices keep oriented along the longitudinal direction. Such a conclusion can be drawn because magnetic sublattices are separately considered and linked by means of an inter-lattice exchange. Not all of its capabilities have been here developed, since, for example, the parameters used for both sublattices are common and only differ in their respective saturation magnetizations and gyromagnetic ratios. Other differences have been already suggested in the literature, such as Gilbert damping,\cite{Seib:10,Ryu:13} or the effect of spin-orbit torques,\cite{Dieny:94, Jiang:06} which can also be covered by our models and could be essential to describe further experimental observations. Indeed, our approach permits to analyze the application of very distinct stimuli acting differently on the sublattices, such as spin transfer torques, both adiabatic and non-adiabatic, the mentioned spin-orbit torques or in-plane magnetic fields. The effects of imperfections and geometry have indeed been addressed. Our models are not only applicable to FiMs deposited on HMs, but to any system that can be described by means of two interlaced subsystems. In fact, due to its versatility, our models can serve as a cornerstone of future experiments regarding CDDWD, which can be of relevance in the development of FiM based devices.

\section*{Acknowledgments}
The authors want to thank Dr. Luis S{\'a}nchez-Tejerina for his valuable comments, and also acknowledge the support by project MAT2017-87072-C4-1-P 
from the Spanish government, and project SA299P18 from the Junta de Castilla y Le\'{o}n.

\appendix
\section{Development of the collective coordinates model (1DM)}\label{appendix:1DM}
The LLG equation (\ref{eqn:01}) also admits a resolution with the help of variational principles, then requiring the description of the magnetization textures by means of some collective coordinates as the instantaneous position $q$ of a DW in the system, and the orientations $\psi_i$ of the in-plane components of the DW moments of each sublattice with respect to the longitudinal axis of the strip. These coordinates determine the local orientation of the magnetization through the well-known ansatz $\theta_i=2\arctan{\text{e}^{Q_i\frac{x-q}{\Delta}}}$ and $\phi_i=\psi_i$, $\theta_i$ and $\phi_i$ defining the local orientation of the magnetization in each sublattice, $\Delta$ accounting for the DW width, and $Q_i$ determining the magnetization transition, as they have been defined in the main text. Previous to the application of the ansatz, system energies can be expressed as:
\begin{subequations}
\begin{align}
\epsilon_{exch,i}&=A_i\left(\nabla\vec{m}_i\right)^2=A_i\left[\left(\nabla\theta_i\right)^2+\sin^2\theta_i\left(\nabla\phi_i\right)^2\right]\text{,}\\
\epsilon_{ani,i}&=K_{u,i}\left[1-\left(\vec{m}\cdot\vec{u}_k\right)^2\right]=K_{u,i}\sin^2\theta_i\text{,}\\
\epsilon_{dmg,i}&=-\frac{1}{2}\mu_0M_{s,i}\vec{H}_{dmg}\cdot\vec{m}_i=\left(K_{dmg,i}+K_{sh,i}\sin^2\phi_i\right)\sin^2\theta_i\text{,}\\
\begin{split}
\epsilon_{ext,i}&=-\mu_0M_{s,i}\vec{H}_{ext}\cdot\vec{m}_i=\\
&=-\mu_0M_{s,i}\left(H_x\sin\theta_i\cos\phi_i+H_y\sin\theta_i\sin\phi_i+H_z\cos\theta_i\right)\text{,}
\end{split}\\
\begin{split}
\epsilon_{DM,i}&=D_i\left[m_{z,i}\left(\nabla\vec{m}_i\right)-\left(\vec{m}_i\cdot\nabla\right)m_{z,i}\right]=\\
&=D_i\left[\cos\phi_i\frac{\partial\theta_i}{\partial x}+\sin\phi_i\frac{\partial\theta_i}{\partial y}+\sin\theta_i\cos\theta_i\left(\sin\phi_i\frac{\partial\phi_i}{\partial x}-\cos\phi_i\frac{\partial\phi_i}{\partial y}\right)\right]\text{,}
\end{split}
\end{align}
\end{subequations}
with $A_i$, $K_{u,i}$, $D_i$ being respectively the intra-sublattice exchange constants, the magnetic uniaxial anisotropy constants, and the anisotropic exchange constants. The magnetostatic interactions $\epsilon_{dmg,i}$ are accounted for by means of two terms: $K_{dmg,i}$ and $K_{sh,i}$, the former entering in the computation of the DW width $\Delta$, and the latter determining the DW type, either Bloch or N{\'e}el type. Due to the low net magnetization of the system, these terms can be considered as negligible if compared with the uniaxial anisotropy and the anisotropic exchange terms, but they are here included for the sake of a more general treatment of the system. Finally, $\left(H_x,H_y,H_z\right)$ are the vector components of any external field $\vec{H}_{ext}$. In addition to these terms, and inter-sublattice exchange interaction must be considered, which is postulated to be in the form: 
\begin{equation}
\epsilon_{12}=-B_{12}{\vec{m}}_1\cdot{\vec{m}}_2=-B_{12}\sin{\theta_1}\sin{\theta_2}\cos{\left(\phi_1-\phi_2\right)}+\cos{\theta_1}\cos{\theta_2}\text{,}
\end{equation}
where the constant $B_{12}$ determines that the interaction promotes antiparallel (parallel) alignment of the sublattices if $B_{12}>0$ ($B_{12}<0$). Field-like spin-orbit torque (FL-SOT) can also be included at this point, and is written as:
\begin{equation}
\epsilon_{SOT,i}=-\mu_0M_{s,i}H_{FL,i}\vec\sigma\cdot\vec{m_i}\text{,}
\end{equation}
where $\vec\sigma=\vec{u}_J\times\vec{u}_z=-\vec{u}_y$, for a longitudinal current so that $\vec{u}_J=\vec{u}_x$,\cite{Ryu:13,Haazen:13,Emori:13} and $H_{FL,i}$ represents the amplitude of the FL-SOT. Each term is to be added to the respective sublattice.

The use of the mentioned ansatz results in the following identities:
\begin{subequations}
\begin{eqnarray}
\nabla\theta_i=\frac{\partial\theta_i}{\partial x}=+Q_i\frac{\sin\theta_i}{\Delta}&\text{,}&\quad\nabla\phi_i=\nabla\psi_i=0\text{,}\\
\delta\theta_i=-Q_i\frac{\sin\theta_i}{\Delta}\delta q&\text{,}&\quad\delta\phi_i=\delta\psi_i=0\text{,}\\
\dot\theta_i=-Q_i\frac{\sin\theta_i}{\Delta}\dot{q}&\text{,}&\quad\dot\phi_i=\dot\psi_i\text{.}
\end{eqnarray}
\end{subequations}
Once all energy density terms have been established, they can be integrated along the longitudinal direction to obtain the total areal energy density $\sigma$ of the system as $\sigma=\int_{-\infty}^{+\infty}\left(\epsilon_1+\epsilon_2 +\epsilon_{12}\right)dx$, resulting in the value:
\begin{equation}
\begin{split}
\sigma&=\frac{2A_1}{\Delta}+2\Delta\left(K_{eff,1}+K_{sh,1}\sin^2{\psi_1}\right)+\pi\ Q_1D_1\cos{\psi_1}-\\
&-\mu_0M_{s,i}\pi\Delta\left(H_x\cos{\psi_1}+H_y\sin{\psi_1}\right)-2Q_1\mu_0M_{s,1}qH_z+\\
&+\frac{2A_2}{\Delta}+2\Delta\left(K_{eff,2}+K_{sh,2}\sin^2{\psi_2}\right)+\pi\ Q_2D_2\cos{\psi_2}-\\
&-\mu_0M_{s,2}\pi\Delta\left(H_x\cos{\psi_2}+H_y\sin{\psi_2}\right)-2Q_2\mu_0M_{s,2}qH_z-\\
&-2B_{12}\Delta\cos\left(\psi_2-\psi_1\right)+\mu_0\pi\Delta\left(M_{s,1}H_{FL,1}\sin\psi_1+M_{s,2}H_{FL,2}\sin\psi_2\right)\text{.}
\end{split}
\end{equation}
This expression has been obtained by using the following intermediate identities: $\int_{-\infty}^{+\infty}\sin\theta_idx=\pi\Delta$, $\int_{-\infty}^{+\infty}\sin^2\theta_idx=2\Delta$, and $\int_{-\infty}^{+\infty}\cos\theta_idx=2Q_iq$.

Additionally to the areal energy term, the kinetic $K_i$ and dissipation $F_i$ terms per unit area for each sublattice are also worked out by integrating along the longitudinal axis their respective expressions:

\begin{subequations}
\begin{align}
\begin{split}
K_i&=\int_{-\infty}^{+\infty}\left[\frac{\mu_0M_{s,i}}{\gamma_i}\phi_i\sin\theta_i\dot\theta_i-\frac{\mu_0M_{s,i}}{\gamma_i}\phi_i\sin\theta_i\left(u_i\vec{u}_J\cdot\nabla\right)\theta_i\right]dx=\\
&=-2Q_i\frac{\mu_0M_{s,i}}{\gamma_i}\left(\dot{q}+u_i\right)\psi_i\text{,}
\end{split}\\
\begin{split}
F_i&=\int_{-\infty}^{+\infty}\frac{\alpha_i\mu_0M_{s,i}}{\gamma_i}\left\lbrace\left[\frac{d}{dt}-\frac{\beta_i}{\alpha_i}\left(u_i\vec{u}_J\right)\right]\vec{m}_i-\frac{\gamma_i}{\alpha_i}H_{SL,i}\vec{m}_i\times\vec{\sigma}\right\rbrace^2dx=\\
&=\frac{\alpha_i\mu_0M_{s,i}}{\gamma_i}\Delta\left[\frac{\dot{q}^2}{\Delta^2}+\dot{\psi}_i^2\right]+\frac{\mu_0M_{s,i}}{\gamma_i}2\beta_iu_i\frac{\dot{q}}{\Delta}+Q_i\mu_0\pi M_{s,i}H_{SL,i}\dot{q}\text{,}
\end{split}
\end{align}
\end{subequations}
It is to note that only the relevant terms are written down in the expression of the dissipation term, i.e., those depending on $\dot{q}$ or $\dot{\psi}_i$. These expressions include the adiabatic spin transfer torques, given by some $u_i$ values,\cite{Thiaville:04} their non-adiabatic counterparts, related to the latter by the $\beta_i$ parameters, and the Slonczewski-like terms of the spin-orbit torques (SL-SOT), accounted for by the $H_{SL,i}$ factors, as they have been defined in the main text. Together with the previously presented intermediate results, the following ones must be considered in this calculation: $\int_{-\infty}^{+\infty}\sin\theta_i\dot\theta_idx=-2Q_i\dot{q}_i$, and $\int_{-\infty}^{+\infty}\dot\theta_i^2dx=\frac{q_i^2}{\Delta}+\Delta\dot\psi_i^2$.

Dynamic equations can be derived by building up the lagrangian $L=\sigma+K$, with $K=K_1+K_2$, and the dissipation function $F=F_1+F_2$, and applying the Euler-Lagrange-Rayleigh conditions $\frac{\partial L}{\partial X}-\frac{d}{dt}\left(\frac{\partial L}{\partial\dot{X}}\right)+\frac{\partial F}{\partial\dot{X}}=0$, with $X=\left\{q,\psi_1,\psi_2\right\}$,\cite{Thiaville:04} resulting in the set of equations presented in the main text. 

\section*{References}
\bibliographystyle{unsrt}
\bibliography{\jobname}

\end{document}